\documentclass{PoS}

\title{The Transform between the space of observed values and the space
of possible values of the parameter}

\author{\speaker{Sergey I. Bityukov}\thanks{This work has been 
supported by grants RFBR 07-02-00256-a.}\\
Institute for high energy physics, 142281 Protvino, Russia\\  
Email: \email{Serguei.Bitioukov@cern.ch}
}

\author{Nikolai V. Krasnikov\\
Institute for nuclear research RAS, Moscow, Russia\\  
E-mail: \email{Nikolai.Krasnikov@cern.ch}
}

\author{Vera V. Smirnova\\
 Institute for high energy physics, 142281 Protvino, Russia\\
E-mail: \email{Vera.Smirnova@ihep.ru}
}
 
\author{Vera A. Taperechkina\\
Moscow State University of instrument engineering and computer science, Moscow, Russia}

\abstract{
In ref~\cite{Bit2004A} the notion of statistically dual distributions is 
introduced. The reconstruction of confidence density~\cite{Bit2005} 
for the location parameter for several pairs of statistically dual 
distributions (Poisson and Gamma, normal and normal, Cauchy and Cauchy, 
Laplace and Laplace) in the case of single observation of the random 
variable is a unique. It allows to introduce the Transform between the 
space of observed values and the space of possible values of the parameter.}

\FullConference{XI International Workshop on Advanced Computing and Analysis Techniques in Physics Research\\
		 April 23-27 2007\\
		 Amsterdam, the Netherlands}

\ShortTitle{The Transform between the space of}

\begin{document}
\tableofcontents

\section{Introduction}\label{sec:intr}

As it is shown in refs.~\cite{Bit2000,Bit2002}, in the framework of 
frequentist approach we can construct the probability distribution 
of the possible magnitudes of the Poisson distribution parameter
to give the observed number of events $\hat n$ in a Poisson stream of events.  
This distribution, which can be called a confidence density function
of a parameter, is described by a Gamma-distribution  
with the probability density function which looks like Poisson distribution 
of probabilities.  This is the reason for naming this pair
of distributions as statistically dual distributions.
Also, the interrelation between the Poisson and Gamma distributions allows 
to reconstruct the confidence density of the Poisson distribution parameter 
in a unique way~\cite{Bit2004A} and, correspondingly, to construct any 
confidence interval for the parameter~\footnote{If we have the procedure 
which states the one-to-one conformity between the observed value of random 
variable and the confidence interval of any level of significance then we can 
reconstruct the confidence density of the parameter in single way. 
Confidence density contains more information than confidence interval.}.

According to B.~Efron~\cite{Efr1998} the confidence
density is the fiducial~\cite{Fis1930} distribution of the parameter. 
This distribution is considered as a genuine a posterior density for 
the parameter without prior assumptions. More detail the confidence 
distributions is considered in ref.~\cite{Sin2005}.

The same relation, which allows one to reconstruct the confidence density of 
a parameter in a unique way, exists between several pairs of statistically 
self-dual distributions (normal and normal, Laplace and Laplace and 
Cauchy and Cauchy). As consequence, the Transform between the space of 
realizations of the random variable and the space of possible values of 
the parameter takes place~\cite{Bit2005, Bit2006} in this case.

Note that the posterior distribution of the parameter is also used for the 
definition of conjugate families in the Bayesian approach.  The interrelation 
between the statistically dual distributions and conjugate families is 
discussed in ref.~\cite{Bit2005}. 

The notion ``statistically dual distributions'' is introduced in the next 
Section. Section~3 describes the Transform between the space of realizations 
of the random variable and the space of possible values of the parameter. 
The method of the confidence density construction for signal
with known background is shown in Section 4 as an example of 
the Transform application. In Section 5 the confidence density  
and the Bayes' Theorem are used for estimation of the uncertainty 
in distinguishing of two simple hypotheses under the experiment 
planning~\cite{Bit2004}. 

\section{Statistically dual distributions}\label{SDD}

Let us define statistically dual distributions.

\underline{Definition 1}: 
Let $\phi (x,\theta)$ be a function of two variables. If the same function 
can be considered both as a family of the probability density functions
(pdf) $f(x|\theta)$
of the random variable $x$ with parameter $\theta$ and as another 
family of pdf's $\tilde f(\theta|x)$ of the random variable $\theta$ with 
parameter $x$ (i.e. $\phi (x,\theta)=f(x|\theta)=\tilde f(\theta|x)$), 
then this pair of families of distributions can be named as 
{\bf statistically dual distributions}.

The  statistical duality of Poisson and Gamma-distributions follows from 
simple discourse.

Let us consider the Gamma-distribution $\Gamma_{1,n+1}$ with probability 
density~\cite{Bit2004}

\begin{equation}
g_n(\mu) = \displaystyle \frac{\mu^n}{n!} e^{-\mu},~ 
\mu > 0,~n > -1.  
\label{eq:1}
\end{equation}

It is a common supposition that the probability of observing $n$ events in the 
experiment 
is described by Poisson distribution with parameter $\mu$, i.e.

\begin{equation}
f(n|\mu)  = \frac{{\mu}^n}{n!} e^{-\mu},~\mu > 0,~n \ge 0.
\label{eq:2}
\end{equation}

One can see that if the parameter and variable in Eq.~(\ref{eq:1})
and Eq.~(\ref{eq:2}) are exchanged, in other respects the formulae are 
identical. As a result these distributions (Gamma and Poisson) 
are {\bf statistically dual distributions}. 

These distributions are connected by the
identity~\cite{Bit2000}~(see, also, this identity in another form in 
refs.~\cite{Jay1976,Fro1979,Cou1995})

\begin{equation}
\displaystyle
\sum_{i = \hat n + 1}^{\infty}{f(i| \mu_1)} +
\int_{\mu_1}^{\mu_2}{g_{\hat n}(\mu) d\mu} + 
\sum_{i = 0}^{\hat n}{f(i|\mu_2)} = 1,~
\label{eq:3}
\end{equation}

\noindent
i.e.

\begin{center}
$\displaystyle
\sum_{i = \hat n + 1}^{\infty}{\frac{\mu_1^ie^{-\mu_1}}{i!}} +
\int_{\mu_1}^{\mu_2}
{\frac{\mu^{\hat n}e^{-\mu}}{\hat n!}d\mu}
+ \sum_{i = 0}^{\hat n}{\frac{\mu_2^ie^{-\mu_2}}{i!}} = 1~$
\end{center} 

\noindent
for any real $\mu_1 \ge 0$ and $\mu_2 \ge 0$ and non-negative integer 
$\hat n$. We can suppose that $\hat n$ is a number of observed events.

The definition of the confidence interval $(\mu_1,\mu_2)$
for the Poisson distribution parameter $\mu$ using~\cite{Bit2000,Bit2005}

\begin{equation}
{P(\mu_1 \le \mu \le \mu_2|\hat n) = 
P(i \le \hat n|\mu_1) - P(i \le \hat n|\mu_2)},
\label{eq:4}
\end{equation}

\noindent 
where $\displaystyle 
P(i \le \hat n|\mu) = \sum_{i=0}^{\hat n}{\frac{\mu^ie^{-\mu}}{i!}}$,
allows one to show
that a Gamma-distribution $\Gamma_{1,1+\hat n}$ is the probability distribution
of different values of $\mu$ parameter of Poisson distribution under condition 
that the observed value of the number of events is equal to $\hat n$, i.e.
$\Gamma_{1,1+\hat n}$ is {\bf the confidence density} of the parameter $\mu$. 
This definition is consistent with the identity Eq.~(\ref{eq:3}).
Note, if we suppose in Eq.~(\ref{eq:3}) that $\mu_1 = \mu_2$ we have a 
conservation of probability. The right-hand side of Eq.~(\ref{eq:4}) 
determines the frequentist sense of this definition. 

Let us consider, for example, 
the  Cauchy distribution with unknown parameter $\theta$ and 
known parameter $b$. Here we also can exchange the parameter $\theta$ and 
variable $x$ while conserving the same formula of the probability distribution.

The probability density of the Cauchy distribution is

\begin{equation}
\displaystyle C(x|\theta) = \displaystyle \frac{b}{\pi (b^2 + 
(x-\theta)^2)}.
\label{eq:5}
\end{equation}

\noindent
The probability density of 
its statistically dual distribution is also the Cauchy distribution:  

\begin{equation}
\displaystyle \tilde C(\theta|x) = 
\displaystyle \frac{b}{\pi (b^2 + (x-\theta)^2)}. 
\label{eq:6}
\end{equation}

\noindent
In this a way the Cauchy distribution can be named as 
{\bf statistically self-dual distribution}. An identity like 
Eq.~(\ref{eq:3}) also holds in another form,

\begin{equation}
{\displaystyle \int_{-\infty}^{\hat x-\delta_1}{C(x|\hat x)dx} + 
\int_{\hat x-\delta_1}^{\hat x+\delta_2}{\tilde C(\theta|\hat x)d\theta} 
+\displaystyle \int_{\hat x+\delta_2}^{\infty}{C(x|\hat x)dx} = 1},
\label{eq:7}
\end{equation}

\bigskip

\noindent
where {$\hat x$} is the observed value of random variable 
{$x$}, {$\hat x - \delta_1$} and {$\hat x + \delta_2$} 
are bounds of the confidence interval for location parameter 
{$\theta$}, and $\tilde C(\theta|\hat x)$ is the confidence density.

Such identity (\ref{eq:7}) is a property of statistically self-dual 
distributions, namely, Cauchy and Cauchy, normal and normal, Laplace and
Laplace.

\section{The Transform between the space of observed values and
the space of possible values of the parameter}\label{sec:EPD}

It is easy to show that the reconstruction of the confidence density is 
unique if Eqs.~(\ref{eq:3}) or (\ref{eq:7}) 
hold~\cite{Bit2004A,Bit2005}.

As a result we have  the {\bf Transform} 
(both for Poisson-Gamma pair of families of distributions
and for statistically self-dual distributions) 

\begin{equation}
\tilde f(\theta|\hat x) = T_{cd} \hat x
\label{eq:9}
\end{equation}

\noindent
between the space of the realizations $\hat x$ of random variable $x$
(with the probability density $f(x|\theta)$)
and the space of the possible values of the parameter $\theta$
(with the confidence density $\tilde f(\theta|\hat x)$).
Here $f(x|\theta)$ is the probability density of either normal, or Cachy,
or Laplace distribution and $\tilde f(\theta|\hat x)$ is the 
corresponding confidence distribution. 

The Transform Eq.~(\ref{eq:9}) allows one to use statistical inferences 
about the random variable for estimation of an unknown parameter.

The simplest examples of this are given by several
infinitely divisible distributions.

\underline{Definition 2}:  
A distribution $F$ is {\bf infinitely divisible} if for
each $n$ there exist a distribution function $F_n$
such that $F$ is the $n$-fold convolution of $F_n$.

As it is known the Poisson, Gamma-, normal and Cauchy distributions are
infinitely divisible distributions.  The sum of independent and identically
distributed random variables, which obey one of the families of 
distributions above, also obeys the distribution from the same family.
Application of the Transform Eq.~(\ref{eq:9})
to this sum allows one to reconstruct the confidence density of the
parameter in the case of several observation of the same random variable.
It means that we construct the relation

\begin{equation}
\tilde f(n\theta|\hat x_1 + \hat x_2 + \dots + \hat x_n) = 
T_{cd} (\hat x_1 + \hat x_2 + \dots + \hat x_n),
\label{eq:10}
\end{equation}

\noindent
where $T_{cd}$ is the operator of 
the Transform Eq.~(\ref{eq:9}), the set $\hat x_1,
\hat x_2, \dots, \hat x_n$ are the observed values. Thereafter we
reconstruct the confidence density of $\theta$, i.e.
$\tilde f(\theta|\hat x_1,\hat x_2, \dots, \hat x_n)$.

The use of the confidence density also can be formulated in  
Bayesian framework.

Let us consider, as an example, the Cauchy distribution.  In our approach 
we suppose that the parameter $\theta$ is not a random value and before
the measurement we do not prefer any of values of this parameter, i.e. 
possible values of the parameter have equal probability and a prior
distribution of $\theta$ is $\pi(\theta)=const$.
Suppose we observe $\hat x_1$ and update our prior via 
the Transform Eq.~(\ref{eq:9}) to obtain
$\tilde C(\theta|\hat x_1)$, which is the pdf of the Cauchy distribution.
This becomes our new prior before observing $\hat x_2$.
It is easy to show that in the case of the observing $\hat x_2$ the 
reconstructed confidence density (or our next new prior) 
$\tilde C(2\theta|\hat x_1+\hat x_2)$~\footnote{
As it is known, if $C(x_1|\theta_1,b_1) =  
\frac{b_1}{\pi (b_1^2 + (x_1-\theta_1)^2)}$ and
$C(x_2|\theta_2,b_2) =  
\frac{b_2}{\pi (b_2^2 + (x_2-\theta_2)^2)}$ then 
$C(x_1+x_2|\theta_1+\theta_2,b_1+b_2) = 
\frac{b_1+b_2}{\pi ((b_1+b_2)^2 + ((x_1+x_2)-(\theta_1+\theta_2))^2)}$
with statistically dual distribution
$\tilde C(\theta_1+\theta_2|x_1+x_2,b_1+b_2) =  
\frac{b_1+b_2}{\pi ((b_1+b_2)^2 + ((x_1+x_2)-(\theta_1+\theta_2))^2)}$.
It means that we can reconstruct $\tilde C(\theta|\hat x_1,\hat x_2)$ 
using $\tilde C(2\theta|\hat x_1+\hat x_2,2b)$ (in our case 
$\theta_1=\theta_2=\theta$ and $b_1=b_2=b$).}
also is the pdf of the Cauchy distribution.
By induction this argument extends to sequences of any number of
observations\\

\begin{center}
$\displaystyle \tilde C(n \theta|\hat x_1+\hat x_2+ \dots + \hat x_n) =  
T_{cd} (\hat x_1 + \hat x_2 + \dots + \hat x_n)$,\\
\end{center}

\noindent
i.e. we use the iterative procedure\\

\begin{equation}
\tilde C(\theta|\hat x_1,\hat x_2, \dots, \hat x_{n-1}, \hat x_n) = 
T_{pd} (\tilde C(\theta|\hat x_1, \hat x_2, \dots, \hat x_{n-1}),\hat x_n),
\label{eq:11}
\end{equation}

\noindent
where $T_{pd}$ is the operator of the Transform between
a prior density and a posterior density of the parameter. 

Note that a prior
density here is only the result of direct calculations of probabilities
in frame of the Transform $T_{pd}$ with the usage of the knowledge about 
the law of distribution of the random variable, i.e. we construct the 
confidence density without any suppositions about a prior (``uniform prior'' 
is not a prior density because if $\pi(\theta)=const$ then 
$\int_{-\infty(~or~0)}^{\infty}{\pi(\theta)d\theta} = \infty$). On the other
hand, a prior knowledge about the law of distribution of the parameter in the 
case of the random origin of parameter can be used for construction of the 
confidence density.

\section{The method of confidence density construction for a signal
with known background}
\label{sect:method}

The confidence density is more informative notion than the confidence
interval and gives many advantages in the construction of
the confidence intervals. For example, the Gamma-distribution 
$\Gamma_{1, \hat n+1}$ is the confidence density of the parameter
of Poisson distribution in the case of the $\hat n$ observed events from the
Poisson flow of events~\cite{Bit2000,Bit2002}. It means that we
can reconstruct any confidence intervals (shortest, central, 
with optimal coverage, \dots)
by the direct calculation of the probability density of Gamma-distribution. 
 
The next example illustrates the advantages of the confidence density 
construction. 
Let us consider the Poisson distribution with two components:
the signal component with a parameter $\mu_s$ and background component
with a parameter $\mu_b$, where $\mu_b$ is known.
To construct confidence intervals for the parameter $\mu_s$ of a signal 
in the case of observed value $\hat n$, we must find the confidence density 
$P(\mu_s|\hat n)$. 

Firstly let us consider the simplest case $\hat n = \hat s + \hat b = 1$.
Here $\hat s$ is the number of signal events and $\hat b$ is the number of
background events among the observed number $\hat n$ of events.

$\hat b$ can be equal to 0 and 1.
We know that $\hat b$ is equal to 0 with probability (Eq.(2))

\begin{equation}
p_0 = f(\hat b = 0|\mu_b) = 
\displaystyle \frac{\mu_b^0}{0!} e^{-\mu_b} = e^{-\mu_b} 
\end{equation}

and $\hat b$ is equal to 1 with probability 

\begin{equation}
p_1 = f(\hat b = 1|\mu_b) = 
\displaystyle \frac{\mu_b^1}{1!} e^{-\mu_b} = 
\mu_b e^{-\mu_b}.  
\end{equation}

Correspondingly, 
$P(\hat b = 0|\hat n = 1) = P(\hat s = 1|\hat n = 1) =
\displaystyle \frac{p_0}{p_0 + p_1}$ and

$P(\hat b = 1|\hat n = 1) = P(\hat s = 0|\hat n = 1) =
\displaystyle \frac{p_1}{p_0 + p_1}$.

It means that the distribution of the confidence density $P(\mu_s|\hat n = 1)$ 
is equal to the sum of distributions

\begin{equation}
P(\hat s=1|\hat n=1) \Gamma_{1,2} + P(\hat s=0|\hat n=1) \Gamma_{1,1} = 
\displaystyle \frac{p_0}{p_0+p_1} \Gamma_{1,2} +
\displaystyle \frac{p_1}{p_0+p_1} \Gamma_{1,1}, 
\end{equation}

\noindent
where $\Gamma_{1,1}$ is the Gamma distribution with the probability
density $g_{\hat s=0}(\mu_s) = \displaystyle e^{-\mu_s}$ 
and $\Gamma_{1,2}$ is the Gamma distribution with the probability
density $g_{\hat s=1}(\mu_s) = \displaystyle \mu_s e^{-\mu_s}$.
As a result, we have the confidence density of the parameter $\mu_s$

\begin{equation}
P(\mu_s|\hat n=1) = 
\displaystyle \frac{\mu_s + \mu_b}{1 + \mu_b} 
\displaystyle e^{-\mu_s}. 
\end{equation}

Using the formula (Eq.(15)) for $P(\mu_s|\hat n = 1)$ and formula (Eq.(4)),
we construct the shortest confidence interval
of any confidence level in a trivial way.

In this manner we can construct the confidence density $P(\mu_s|\hat n)$
for any values of $\hat n$ and $\mu_b$. We have obtained 
the known formula~\cite{7,8,9}

\begin{equation}
P(\mu_s|\hat n) = \displaystyle 
\frac{(\mu_s+\mu_b)^{\hat n}}{\hat n! \displaystyle \sum_{i=0}^{\hat n}{\mu_b^i \over {i!}}}
\displaystyle e^{-\mu_s}. 
\end{equation}

The numerical results for the
confidence intervals 
are shown in Table~1.

\begin{table}
\begin{tabular}{r|cccccccccc} \hline
~$\hat n\backslash\mu_b$~  & 
         0.0   &     1.0    &     2.0   &    6.0  &   12.0      &  15.0  \\
\hline
0 & 0.00, 2.30 & 0.00, 2.30& 0.00, 2.30&0.00, 2.30&0.00, 2.30 & 0.00, 2.30\\  
1 & 0.09, 3.93 & 0.00, 3.27& 0.00, 3.00&0.00, 2.63&0.00, 2.48 & 0.00, 2.45\\  
2 & 0.44, 5.48 & 0.00, 4.44& 0.00, 3.88&0.00, 3.01&0.00, 2.68 & 0.00, 2.61\\  
3 & 0.93, 6.94 & 0.00, 5.71& 0.00, 4.93&0.00, 3.48&0.00, 2.91 & 0.00, 2.78\\  
4 & 1.51, 8.36 & 0.51, 7.29& 0.00, 6.09&0.00, 4.04&0.00, 3.16 & 0.00, 2.98\\  
5 & 2.12, 9.71 & 1.15, 8.73& 0.20, 7.47&0.00, 4.71&0.00, 3.46 & 0.00, 3.20\\ 
6 & 2.78,11.05 & 1.79,10.07& 0.83, 9.01&0.00, 5.49&0.00, 3.80 & 0.00, 3.46\\  
7 & 3.47,12.38 & 2.47,11.38& 1.49,10.37&0.00, 6.38&0.00, 4.19 & 0.00, 3.74\\  
8 & 4.16,13.65 & 3.18,12.68& 2.20,11.69&0.00, 7.35&0.00, 4.64 & 0.00, 4.06\\  
9 & 4.91,14.95 & 3.91,13.96& 2.90,12.94&0.00, 8.41&0.00, 5.15 & 0.00, 4.42\\  
10 &5.64,16.21 & 4.66,15.22& 3.66,14.22&0.02, 9.53&0.00, 5.73 & 0.00, 4.83\\ 
20 &13.50,28.33&12.53,27.34&11.53,26.34&7.53,22.34&1.70,16.08 & 0.00,12.31\\
\end{tabular}
\caption{90\% C.L. intervals for the Poisson signal mean $\mu_s$, 
for total events observed $\hat n$, for known mean background $\mu_b$
ranging from 0 to 15.}
\label{tab1}
\end{table}

\section{The ``Inverse Transform''}

In this Section the approach to estimation of quality of planned 
experiments~\cite{Bit2004} is used to show the possibility of the 
``Inverse Transform''. This approach is based on the 
analysis of uncertainty, which will take place under the future hypotheses 
testing about the existence of a new phenomenon in Nature. 
We consider a simple statistical hypothesis $H_0$: {\it new physics is present 
in Nature} (i.e. $\mu = \mu_s + \mu_b$ in the Eq.(2)) against a simple 
alternative hypothesis $H_1$: {\it new physics is absent} ($\mu = \mu_b$). 
The value of uncertainty is determined by the values of
the probability to reject the hypothesis $H_0$ when it is true 
(Type I error $\alpha$) and the probability to accept the hypothesis $H_0$ 
when the hypothesis $H_1$ is true (Type II error $\beta$). 
This uncertainty characterises the distinguishability of the hypotheses
under the given choice of critical area.

Let both values $\mu_s$ and $\mu_b$, which are defined in the previous
Section, be exactly known. 
In this simplest case the errors of Type I and II, which will take place in 
testing of hypothesis $H_0$ versus hypothesis $H_1$, can be written 
as follows:

\begin{equation}
\cases{ 
\displaystyle
\alpha=\sum^{n_c}_{i=0}{f(i|\mu_s+\mu_b)}, \cr
\displaystyle
\beta=1-\sum^{n_c}_{i=0}{f(i|\mu_b)},
}
\end{equation}

\noindent
where $n_c$ is a critical value and $f(i|\mu)$ is defined by the Eq.(2). 

Let the values $\hat \mu_s = \hat s$ and 
$\hat \mu_b = \hat b$ be known, for example, from Monte Carlo 
experiment with integral luminosity which is exactly the same as the
data luminosity later in the planned experiment. It means that we must
include the uncertainties in values $\mu_s$ and $\mu_b$ to the system
of the equations Eqs.(17). As it is shown in ref.~\cite{Bit2004} 
(see, also, the generalised case in the same reference) we have the system

\begin{equation}
\cases{
\displaystyle \alpha = 
\int_{0}^{\infty}{g_{\hat s+ \hat b}(\lambda)\sum_{i=0}^{n_c}f(i|\lambda)d\lambda} 
= \sum_{i=0}^{n_c} \frac{C^i_{\hat s+ \hat b+i}}{2^{\hat s + \hat b+i+1}}, \cr
\displaystyle \beta = 1 -
\int_{0}^{\infty}{g_{\hat b}(\lambda)\sum_{i=0}^{n_c}f(i| \lambda)d\lambda}
= 1 - \sum_{i=0}^{n_c} \frac{C^i_{\hat b+i}}{2^{\hat b+i+1}},
}
\end{equation}

\noindent
where the critical value $n_c$ under the future 
hypotheses testing about the observability can be chosen in accordance with
the test of equal probability~\cite{Bit2000B} and 
 $C^i_N$ is $\displaystyle \frac{N!}{i!(N-i)!}$.

Note, here the Poisson distribution is a prior distribution of the
expected probabilities and the negative binomial (Pascal)
distribution is a posterior distribution of the expected probabilities
of the random variable. This transformation of the estimated
confidence densities $g_{\hat s+ \hat b}(\lambda)$ and $g_{\hat b}(\lambda)$
(probability densities of the corresponding $\Gamma-$distributions)
to the space of the expected values of the random variable can be
named the ``Inverse Transform''.

\section{Conclusions}

We have shown that the statistical duality allows one to connect the 
estimation
of the parameter with the measurement of the random variable of the
distribution due to the Transform Eq.~(\ref{eq:9}). 
It gives the tool to construct the confidence densities.

The usage of the confidence densities for the construction of the
confidence intervals and for the construction of a posterior distributions
of probabilities is presented in examples.

\acknowledgments

The authors are grateful to V.B.~Gavrilov, V.A.~Kachanov, Louis Lyons, 
V.A.~Matveev and V.F.~Obraztsov for the interest and useful comments, 
Yu.P.~Gouz, A.~Nikitenko and C.~Wulz for fruitful discussions and 
E.A.~Medvedeva for help in preparing the paper. 


\end{document}